\documentclass[
  aip,
   jcp,
  reprint,           
  superscriptaddress
  nofootinbib,
  longbibliography,
  numerical
]{revtex4-2}

\usepackage{amsmath} 
\usepackage{amssymb} 
\usepackage{graphicx}
\usepackage[hidelinks]{hyperref}
\hypersetup{
    colorlinks=true,
    linkcolor=blue,
    filecolor=magenta,      
    urlcolor=blue,
    citecolor=blue,
}
\usepackage{derivative}
\usepackage{cleveref}
\usepackage{newtxtext} %
\usepackage{newtxmath} %
\usepackage{booktabs} %
\usepackage{caption}  %
\usepackage{siunitx}
\usepackage{microtype}
\usepackage{soul}

\soulregister\cref7
\soulregister\Cref7
\soulregister\cite7
\soulregister\namecref7
\soulregister\labelcref7

\renewcommand{\hl}[1]{#1}

\graphicspath{ {./figs/} } 

\newcommand{\sfig}[4][\linewidth]{
    \begin{figure}[htbp] %
        \centering
        \includegraphics[width=#1]{#2}
        \caption{#3}
        \label{#4}%
    \end{figure}%
  }%
\newcommand{\rv}{\mathbf{r}}
\newcommand{\xv}{\mathbf{x}}
\newcommand{\nvac}{n_\mathrm{vac}}

\newcommand{\cry}{\mathrm{cr}}
\newcommand{\ctwo}{c^{(2)}}
\newcommand{\cone}{c^{(1)}}

\newcommand{\tb}{}
\newcommand{\tr}{}

\usepackage[normalem]{ulem}

\begin{document}

\title{Density Profiles and Direct Correlation Functions from Density Functional Theory in Binary Hard-Sphere Crystals: Substitutional Solid and Interstitial Solid Solution}

\author{Alessandro Simon}
\email{alessandro-rodolfo.simon@uni-tuebingen.de}
\affiliation{Institute of Applied Physics, University of T\"ubingen, Germany}

\author{Martin Oettel}
\affiliation{Institute of Applied Physics, University of T\"ubingen, Germany}
\email{martin.oettel@uni-tuebingen.de}

\date{\today}

\begin{abstract}
  We determine the fully resolved equilibrium density profiles for two binary hard-sphere crystal structures using classical density functional theory through the White Bear II functional from fundamental measure theory.
  While for the substitutional crystal, in which some hard spheres are replaced by spheres of slightly smaller diameter, the density profiles are rather similar
  to the single-component case (narrow Gaussian peaks centered at fcc lattice sites), we observe a more complex behavior for the case of interstitial solid solutions, where the small species is fairly delocalized in the unit cell. 
  Further, we compute the species-resolved inhomogeneous two-body direct correlation functions, \hl{depending on two three-dimensional vectors}, for these two types of binary crystals. The large--large components are mainly determined by the vacancy concentration $\nvac$ and show a characteristic magnitude $\sim 1/\nvac$. Based on this observation, we propose a simple geometric picture. The components of the direct correlation function involving the small spheres substantially differ in interstitial solid solutions from those of the substitutional crystal.
\end{abstract}
\maketitle

\section{Introduction}
Hard-sphere (HS) systems are %
ideally suited for understanding the phase behavior of colloidal systems and structural properties of the liquid and the solid phase, owing to the existence of analytical approximations and of extensive numerical studies
\cite{santosbook}. %
A central object of liquid state theory is the family of direct correlation functions (DCFs), which are computed by repeated differentiation of the excess free energy w.r.t.\ the density field. 
While the first-order DCF $\cone(\rv)$ gives information on the non-local chemical potential of the substance, the second order DCF $\ctwo(\rv, \rv')$ is connected to measurable structural
observables like the radial distribution function and the structure factor.
In the \textit{liquid} phase, the Percus--Yevick (PY) solution (for one- and multicomponent systems) gives a satisfactory description of direct correlation functions and it shows that the repulsive core
leads to a negative contribution in the DCF 
, \hl{with a range that is given by the hard-sphere diameter for the one-component system or by the sum of the hard-sphere radii in the case of a multicomponent system.}
The pair correlation function follows from the bulk Ornstein--Zernike (OZ) equation and shows the
characteristic layering behavior found in many liquids. Simulation results for the DCF can be obtained from sampling the pair correlation function and using OZ in the other direction; these results essentially validate the PY solution.

For a one-component HS \textit{crystal} whose position and orientation is fixed in space, both the direct and pair correlation function depend on two position arguments and are connected by the inhomogeneous OZ equation \cite{hansen2013theory} 
\begin{equation}
  \label{eq:oz}
  h(\rv, \rv') = \ctwo(\rv, \rv') +  \int \odif{\rv''} \rho(\rv'')  \ctwo(\rv, \rv'') h(\rv'', \rv'). 
\end{equation}
Here, $\ctwo$ is the DCF while $h=g-1$ and $g$ is the standard pair correlation function, and $\rho(\rv)$ is the density profile in the crystal. In principle, this equation allows the calculation of $\ctwo$ from simulations if $\rho(\rv)$ and $g(\rv, \rv')$ are determined by direct sampling. For HS crystals, $\rho(\rv)$ has been determined in such targeted simulations \cite{oettel2010free}, but to our knowledge there is no simulation work on $g(\rv, \rv')$ for any 3D crystal. On the other hand, both $\rho(\rv)$ and the DCF $\ctwo(\rv, \rv')$ can be obtained from classical density functional theory (DFT). Using the highly accurate White-Bear-II functional \cite{hansen2006density}, this was done in Ref.~\cite{lin2021direct} and gave the surprising result that (i) the crystal DCF is very different in size and range from the liquid and (ii) that it diverges for an ideal (vacancy-free) crystal. This divergence is also reflected in generalized elastic constants that separately measure the free energy response to unit cell deformations and changes of the average density \cite{lin2021direct}. However, even with $\rho$ and $\ctwo$ at hand, no solution of the inhomogeneous OZ \Cref{eq:oz} has been attempted due to its frightening complexity.%

Two-component (binary) HS systems show complex phase behavior\cite{royall2024colloidal}. Crystalline phases include solid solutions, substitutional crystals (SC) for size-similar spheres, stoichiometric alloys (AB, AB$_2$, ...)\cite{eldridge1993superlattice,filion2009prediction}, and Laves phases\cite{hynninen2009stability}.
For certain size ratios and compositions the system will form an interstitial solid solution (ISS) \cite{filion2011self}. This phase is characterized by the larger species forming a face-centered cubic (fcc) lattice and
the smaller species being mainly confined to the (octahedral) interstitials but not necessarily being constrained to a lattice. This is different
from a substitutional binary crystal, where on some lattice sites the bigger species is replaced by the smaller species.%

In this paper, we are interested in the DCFs of the ISS and the SC phase and their relation to the one-component crystals DCF of Ref.~\cite{lin2021direct} as well as the two-component PY solutions for the liquid phase.
We study non-ideal crystals with point defects, i.e. with non-vanishing vacancy densities, which turn out to be larger for the binary crystals than for single component crystals.
As in Ref.~\cite{lin2021direct}, our approach is based on classical density functional theory \cite{evans1979nature,roth2010fundamental,rosenfeld1989,curtin1986density}, which rests on an accurate
free energy functional $F[\rho] = F_\text{id}[\rho] + F_\text{exc}[\rho]$, describing the hard-core interaction. The equilibrium
distribution then follows by minimizing the grand potential $\Omega[\rho] = F[\rho] + \int \odif{\rv} \rho(\rv) \left( V_\text{ext}(\rv) - \mu \right)$
\begin{equation}
  \label{eq:gp}
  \fdv{\Omega[\rho; \mu, T, V]}{\rho(\rv)}_{\rho_\text{eq}(\rv)} = 0.
\end{equation}
Since we are working in a two-component system, the density is a two-component vector, $\rho(\rv)=\{\rho_L(\rv),\rho_S(\rv)\}$, of the density profile for large ($L$) and small ($S$) spheres. Likewise $\mu=\{\mu_L,\mu_S\}$.
The external potential $V_\text{ext}(\rv)$ (also a two-component vector) is set to zero as we are interested in free (unstressed) crystals.

Once the equilibrium density profiles are determined, the direct correlation function $\ctwo$ can then be computed by functional
differentiation of the excess functional
\begin{equation}
      c^{(2)}_{ij}(\rv, \rv') \equiv -\beta \frac{\delta^2 F_{\text{exc}}[\rho]}{\delta \rho_i(\rv) \, \delta \rho_j(\rv')},
      \label{eq:c2ij}
\end{equation}
where the indices $i,j=\{L,S\}$.

\section{Model definitions}
\label{sec:model}

We consider a binary mixture of hard spheres with diameters $\sigma_L$ (for large spheres) and
$\sigma_S = q \sigma_L$ (for small spheres), with a relative size ratio of $q$. The total packing fraction
$\eta$ decomposes into
\begin{equation}
  \label{eq:packing}
  \eta = \eta_L + \eta_S = \eta_L \left(1 + q^3 \frac{x_S}{1-x_S} \right).
\end{equation}
The composition $x_S$ is determined by the ratio of number of small spheres $N_S$ to the number $N_\mathrm{tot}=N_S+N_L$ of small and large spheres,
\begin{equation}
  \label{eq:comp}
  x_S = \frac{N_S}{N_\mathrm{tot}}.
\end{equation}

In the binary mixtures considered, the large species forms an fcc lattice in equilibrium. In the case of the SC, some large spheres can be replaced by small spheres, i.e.\ each fcc lattice site is either occupied by a large or a small sphere, depending on the composition.
In the ISS case, fcc lattice sites are almost exclusively occupied by the large spheres and the smaller spheres predominantly occupy the octahedral holes of the fcc lattice (see also \Cref{fig:vis-iss-sc}). 

\textit{Substitutional crystal:} 
Simulations show that the SC phase is stable for particle-size ratios of $1 > q > 0.85$ \cite{royall2024colloidal,kranendonk1991coexistence}.
In our calculations we chose $q = 0.9$ and $x_S=0.3$. Introducing the relative vacancy concentration $\nvac$ as the probability for a lattice site to be unoccupied or vacant,
we obtain for the unit cell:
\begin{equation}
N_\mathrm{tot} = 4 (1  -  \nvac) = N_L + N_S\;.
\end{equation}
We chose the packing fraction $\eta=0.5511$, which corresponds to a total bulk density $\rho_\text{tot}^\text{bulk} \sigma_L^3=1.1459$. The unit cell length  
\begin{equation}
  \ell = \sqrt[3]{ \frac{4(1-\nvac)}{\rho_\text{tot}^\text{bulk}} },
\end{equation}
is not fixed yet, since it depends on $\nvac$, and this is determined by minimizing the free energy per particle (see below). 
As a side remark, if the crystal consisted
only of large spheres, \hl{the total bulk density would correspond} to a packing fraction of 0.60.

\textit{Interstitial solid:}
Here, simulations show that the ISS phase is starting to become stable at $\eta_L = 0.54, \eta_S > 0$, with an increasing window of stable $\eta_S$ for larger $\eta_L$ (see Figure~5.5 of Ref.~\cite{filion2011thesis} for a phase diagram). The range of stable size ratios is roughly $0.2 < q < 0.42$.
We chose the following parameters: $\eta_L = 0.55$, $x_S = 0.3$ and $q = 0.3$. This leads to a total packing fraction of
$\eta = \eta_L + \eta_S = 0.5564$, which according to simulations \cite{filion2011self} is (narrowly) inside the ISS phase region.
Here we define $\nvac$ as the probability for a lattice site to be unoccupied by a large particle, i.e.\ in the unit cell we have
\begin{equation}
N_L = 4 (1  -  \nvac),
\end{equation}
and consequently the unit cell length is given by
\begin{equation}
  \label{eq:length-def}
  \ell = \sqrt[3]{ \frac{4(1-\nvac)}{\rho_L^\text{bulk}} },
\end{equation}
and, as before, it will be determined by minimizing the free energy per particle with respect to $\nvac$.

\sfig[.8\linewidth]{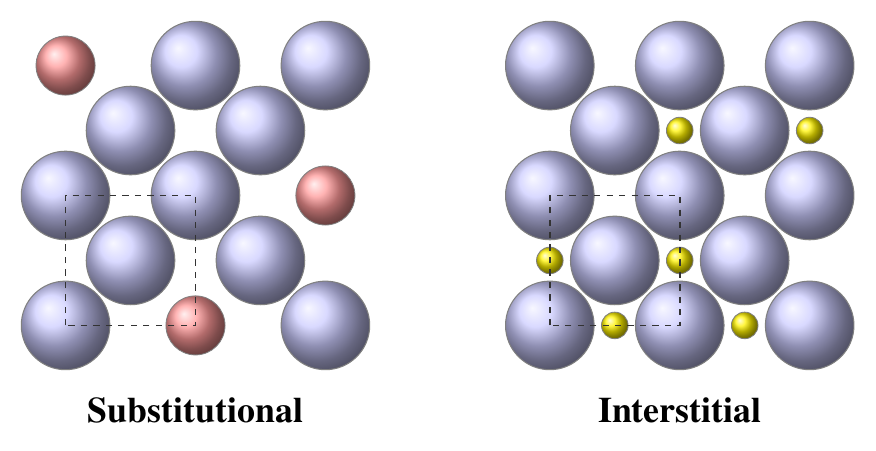}{Sketch of the [100] plane illustrating the substitutional and interstitial solid binary crystal. For the substitutional crystal, the fcc lattice sites are randomly  occupied by large and small particles, while for the interstitial solid structure the fcc lattice is occupied by large particles and additional small particles  predominantly reside in the octahedral voids. In both cases vacancies may also appear. \tb{For the substitutional crystal they correspond to a vacant lattice site while for the interstitial solid they correspond to lattice sites not occupied by a large particle (but small particles may occupy these).}    The dashed line shows one unit cell.}{fig:vis-iss-sc}

\subsection{Fundamental measure theory}
The free energy of a configuration is determined by minimizing the tensorial form of the
White Bear II (WBII-t) functional \cite{hansen2006density} for mixtures. 
The WBII-t functional is very accurate in the description of the solid-- fluid coexistence packing fractions and density profiles of the one-component HS system \cite{oettel2010free} and also of solid--fluid interface tensions and density profiles of the interface \cite{haertel2012prl,oettel2012comparison,oettel2012modes}. It is a success of the FMT tensor functionals that the equilibrium vacancy concentration in the one-component HS crystal comes out small, at coexistence it is $\sim \num{2e-5}$ (simulations give\cite{kwak2008simvac} $\sim 10^{-4}$). Other variants of FMT tensor functionals can also reproduce the equilibrium $\nvac$ from simulations, such as the class of functionals based on Santos' consistent free energy \cite{hansen-goos2015santos} or Lutsko's explicitly stable functionals \cite{lutsko2020explicitly,schoonen2022prl} but show differences to the WBII-t functional with regard to coexistence properties. Since all these functionals share a similar mathematical structure, we do not expect large differences in the properties of the crystalline state and restrict our investigations to the WBII-t functional. Its full definition can be found in the Appendix.

It is sufficient to minimize the free energy in the cubic unit cell which is discretized by $128^3$ points. Numerically, it is nearly impossible to use the conventional minimization scheme of fixing the chemical potential, starting iterations of the minimization \Cref{eq:gp} from a homogeneous starting profile and get the
correct crystal density profile \cite{lutsko2020classical,lutsko2020explicitly}. Rather, we adopt a two-stage minimization by first fixing $\nvac$ (thereby fixing the number of particles in the unit cell and thus also $\ell$) and minimize with respect to $\rho(\rv)$, and secondly minimize with respect to $\nvac$.
Thus the fully minimized free energy is given by
\begin{equation}
\frac{F_\cry}{N_\mathrm{tot}} = \mathrm{min}_{\nvac} \mathrm{min}_{\rho_L(\rv),\rho_S(\rv)} \left. F \left[ \rho_L(\rv), \rho_S(\rv), \nvac \right] / N_\mathrm{tot} \right|_{\nvac}.
\end{equation}
In the first minimization step we initialize the density with Gaussian
density peaks around lattice points, i.e. for the SC these are Gaussian peaks for $\rho_L(\rv)$ and $\rho_S(\rv)$ at the fcc lattice points and
for the ISS these are Gaussian peaks for $\rho_L(\rv)$ at the  fcc lattice points and for $\rho_S(\rv)$ at the (octahedral) interstitials.
During the minimization the density field is free to change its value inside the whole unit cell (only subject to the requirement of constant particle numbers which is ensured by two Lagrangian multipliers $\mu_L', \mu_S' $ which become the equilibrium chemical potentials $\mu_L, \mu_S$ in the fully minimized state).
This is especially important for the small species density field, which will 
spread over the whole unit cell.  The final iteration error of the density profiles (mean squared difference) is of the order \num{1e-6}.

\section{Results}
\label{sec:therm}

\subsection{Density profiles and vacancy concentrations}

For both the substitutional and the interstitial crystal phase there
will be a non-vanishing vacancy density at equilibrium. 
For the one-component crystal, an estimate with Widom's trick gives $\beta \mu' \approx -\ln \nvac + \text{const.} $ \cite{oettel2010free}, i.e.
the Lagrange multiplier ensuring constant particle number in the unit cell weakly diverges for $\nvac \to 0$. 
The divergence seen in the DFT results for $\beta \mu'$ is somewhat weaker, and the equilibrium vacancy concentration close to melting\cite{oettel2010free} is $\nvac \approx \num{2.18e-5}$.

\sfig{batch_subst_eta60/fracL_0.7000/densities_vac_1.0000e-05.pdf}{Equilibrium density profile in the unit cell ([100] plane, $z=0$) for the SC crystal with $q=0.9$, $x_S = 0.3$ and \hl{$\eta=0.5511$}. Panel (a)\ shows the large species while (b)\ shows the smaller species density. }{fig:eq-density-sc}

We expect a similar vacancy concentration for the substitutional binary crystal, since it is based
on the same lattice structure. For the ISS case, higher vacancy densities are observed in
simulations \cite{filion2011self}, especially at packing fractions close to melting.

\textit{Substitutional crystal:}
The equilibrium vacancy concentration for the chosen parameters is determined as $\nvac = $ \num{1e-5}, i.e. only slightly smaller than that for the one-component crystal at coexistence. Note that the total packing fraction of the SC (0.5511) is very close to the WBII-t coexistence packing fraction of 0.544, so this result is not very surprising.    
The density distribution for the binary system at this equilibrium $\nvac $  is shown in \Cref{fig:eq-density-sc}.     
Qualitatively, both peaks show the same form, with the smaller species being slightly more de-localized due to its larger
free volume around the lattice site. The peaks are very isotropic and close to a Gaussian, again very similar to the one-component case.

\sfig[\linewidth]{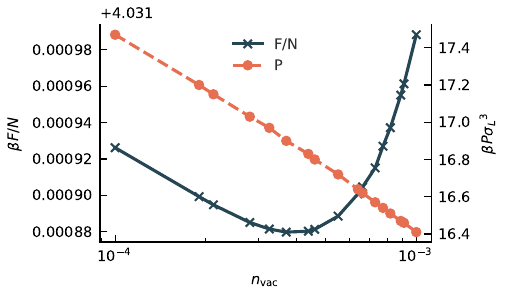}{Free energy per particle ($N \equiv N_\mathrm{tot}$) and pressure as function of $\nvac$ (after the first minimization). Note, that the free energy axis contains an additive offset as indicated at the top of the axis. }{fig:fN_vacancy}

\sfig[\linewidth]{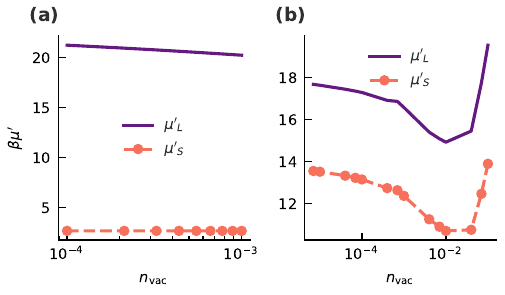}{Lagrange multipliers \hl{interpreted as} canonical chemical potentials $\mu_L'$ and $\mu_S'$  for the ISS case (a) and the SC case (b) at varying vacancy density.}
{fig:mu-comparison}

\textit{Interstitial solid:}
Here, the equilibrium value for the vacancy concentration for the chosen parameters is $\nvac = $ \num{3.7e-4} and is therefore larger than in the one-component crystal case (although again the packing fractions are comparable). 
The dependence of the free energy per particle and the pressure on $\nvac$ is shown in \Cref{fig:fN_vacancy}, where the pressure is defined as
$P=-F/V+\mu_L' \rho_L^\text{bulk} + \mu_S'\rho_S^\text{bulk}$, i.e., the Lagrange multipliers $\mu_L'$ and $\mu_S'$, \hl{ensuring fixed $\nvac$ and $N$},  are interpreted as canonical chemical potentials. It is seen that the free energy only weakly depends on $\nvac$ (note the variation in the fifth digit) but the (canonical) pressure shows a strong dependence. This dependence of $P$ on $\nvac$ essentially originates from the $\mu_L'(\nvac)$ which is monotonically decreasing while $\mu_S'(\nvac)$ is approximately constant, see \Cref{fig:mu-comparison}(a). (Note that for the SC both  $\mu_L'$ and $\mu_S'$ vary strongly with $\nvac$, see \Cref{fig:mu-comparison}(b).)
The equilibrium density profiles in the [100] plane are shown in \Cref{fig:eq-density-iss} (note the logarithmic scale in the density colorbar). 
The large species density shows very isotropic peaks at the fcc lattice sites, similar to the SC and the one-component solid. The small species density shows distorted peaks around the interstitial sites, and a non-vanishing density of $O(10^{-2})$ in the bridging region between two interstitial sites. \tr{At the fcc lattice sites, the density of the small spheres is nonzero, $\approx$ \num{1.85e-3}, showing that small spheres occupy some of the vacant sites of the large sphere lattice. The fraction of the fcc lattice sites occupied by the small spheres can be obtained from the value of the small sphere contribution to the weighted density $n_3$ (signifying a local packing fraction, for the definition see the Appendix), numerically this fraction is $\approx$ \num{3.05e-5}; this value has to be subtracted from $\nvac$ for an estimate of truly vacant fcc lattice sites which is then $\approx$ \num{3.4e-4} . } 
The distorted character of the small species profile becomes also clear in the 3D representation of the total density in the half-unit cell in \Cref{fig:volume-density}. The bright peaks correspond to the Gaussian peak-like contribution of the large species density whereas the small species density leads to the dark, bulky contributions around the interstitial sites.

\sfig{batch_iss_eta55/xs_0.3000/log_densities_vac_3.7000e-04.pdf}{Equilibrium density profiles (large spheres in (a) and small spheres in (b)) in the [100] plane ($z=0$) of the unit cell for the ISS crystal phase \hl{with $q=0.3$, $x_S = 0.3$ and $\eta=0.5564$}.}
{fig:eq-density-iss}


\sfig[1\linewidth]{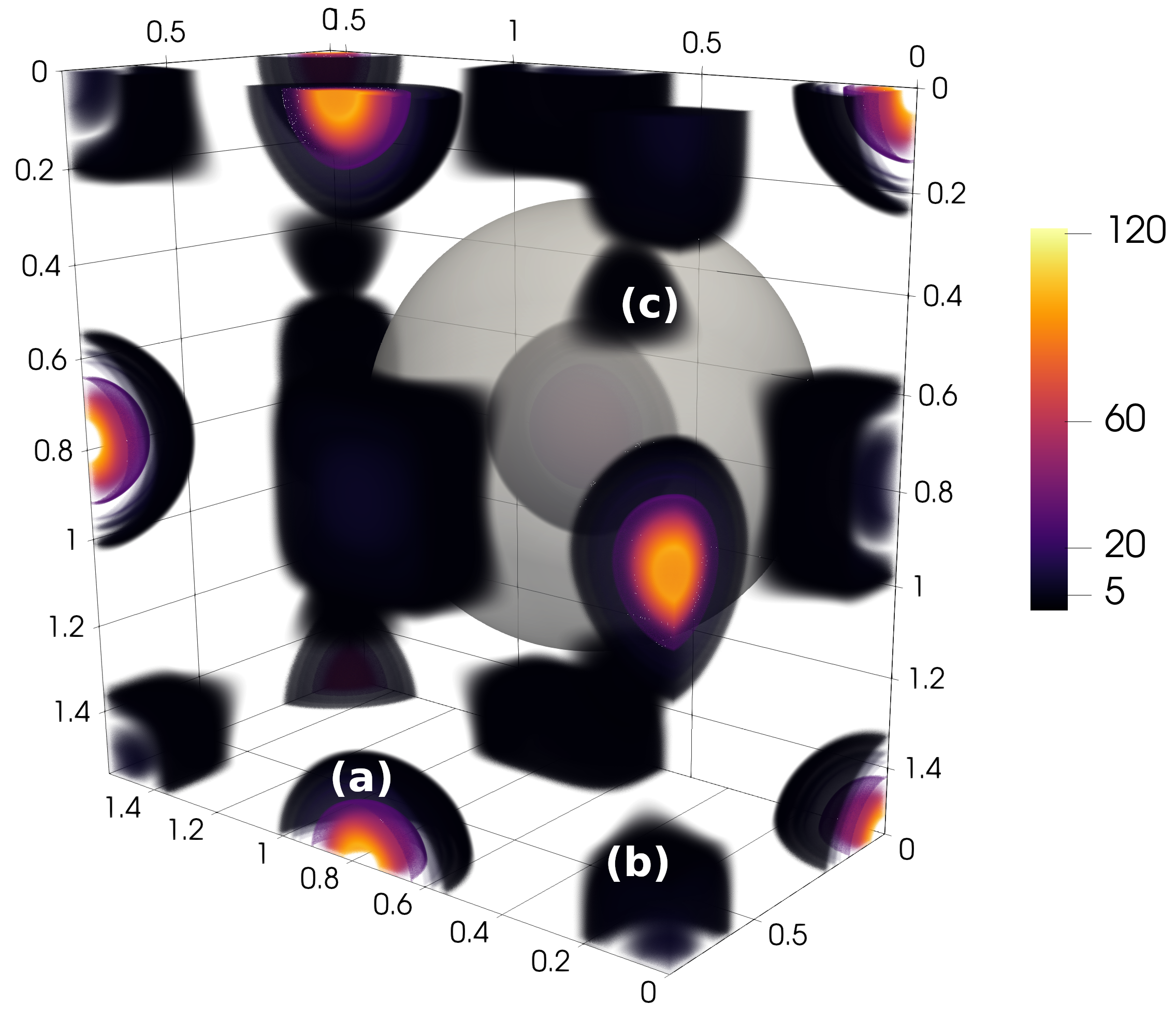}{The volumetric plot shows the total particle density $\rho_L + \rho_S$ in a reduced unit cell (cut in half along one dimension). For reference we also show (in grey) one big-species sphere at its lattice site.
  \hl{Specifically, we show equal density surfaces for representative values of the density. One sees that the total density is high at
    large-sphere lattice sites (a) and much lower at the octahedral interstitials (b) and the tetrahedral interstitials (c). On the other
    hand the density distributions at the interstitials are highly anisotropic. Note for example how the small-species density at the
  octahedral interstitials is clearly influenced by the blocked volume of the large sphere shown as grey sphere.}}
{fig:volume-density}

\subsection{Binary direction correlation function}
\label{sec:DCF}
The species-resolved direct correlation function $c_{ij}^{(2)}$ is defined via the second functional derivative of the excess free energy, see \Cref{eq:c2ij}.
Applying the functional chain rule twice to the FMT form of the excess free energy (see \Cref{eq:Phi}) yields the general FMT form of $\ctwo$, involving the Hessian of the free energy density and the convolution of the geometric weights
\begin{equation}
    c_{ij}^{(2)}(\rv, \rv') = -\beta \sum_{\alpha, \beta} \int d\xv \, 
    \frac{\partial^2 \Phi}{\partial n_\alpha \partial n_\beta}\bigg|_{\{n(\xv)\}} 
    w_\alpha^{(i)}(\xv - \rv) \, w_\beta^{(j)}(\xv - \rv'),
    \label{eq:c2-full}
\end{equation}
\hl{where $\Phi$ and $w^{(i)}_{\alpha}$ are defined in the Appendix.}
From the structure of \Cref{eq:c2-full} we can already deduce the maximal range of the DCF in FMT. It is partly
influenced by the range of the weight functions $w$ and by the support of the second derivative %
of the free energy.
\hl{
The maximal the range of the DCF follows from the convolutional structure
in \Cref{eq:c2-full} and is equal to $r_\text{max} = \sigma_i/2 + \sigma_j/2$ for the component $c_{ij}^{(2)}$.
}

In contrast to the liquid case, the DCF for the crystal is a function in six-dimensional space. Further, it is symmetric under discrete lattice symmetries in the first and second index ($\rv, \rv'$) respectively. %
In the following, we call $\rv$ the reference point and compute the DCF as a function of the shifted second argument $\Delta \rv=\rv'-\rv$, and show it either along  one of the crystallographic directions $[100], [110] \text{ and } [111]$ or as a 3D plot. 
Since we are treating a mixture of particles, the DCF becomes
a matrix in species indices $i, j$.

\sfig[\linewidth]{batch_subst_eta60/fracL_0.7000/densities_vac_1.0000e-05_DCF_c2_100.pdf}{The DCF of the substitutional crystal with \hl{$q=0.9$, $x_S = 0.3$ and $\eta=0.5511$} along the [100] direction evaluated at the fcc site, where $d = |\rv - \rv'|$. } {fig:subst-c2-100}

\subsubsection{Substitutional crystal}
Here, we pick as the reference point an fcc lattice site and show the DCF along the [100] direction
in \Cref{fig:subst-c2-100}. Since the hard spheres in the mixture are of similar size, we expect minor
differences between the species-resolved components. Indeed, we observe that they are all quite similar, with the smaller spheres having slightly smaller range than between the large spheres.
Apart from this the general form is very similar to the single crystal case \cite{lin2021direct}. In particular, the overall magnitude is 
$-O(1/\nvac)$
and therefore orders of magnitude larger than for a typical liquid and there is a pronounced minimum for $|\rv - \rv'|$ near $\sigma_L/2$. The range of the DCF is considerably shorter than the upper bound $r_\text{max}$ derived from the general formula \eqref{eq:c2-full}.

\sfig[0.8\linewidth]{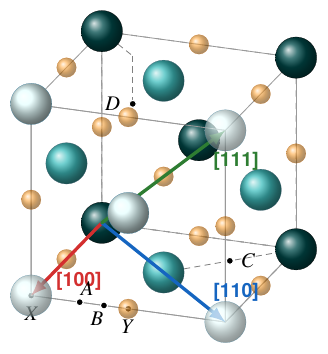}{Sketch of the ISS unit cell and the three crystallographic directions along which we plot the DCF. The smaller orange spheres sit on the interstitial sites, while the large spheres sit on the fcc grid, and their color encodes their distance to the view plane. The labels refer to the reference points described in the main text.}{fig:3d-directions}

\subsubsection{Interstitial solid}

In order to systematically visualize the DCF we chose the following presentation method. First, we vary the reference point
of the first density variation, i.e.\ the variable $\rv$ in \Cref{eq:c2-full}.
We concentrate on the following reference points, and provide their coordinates in the unit cell (there are of course multiple equivalent points):
\begin{itemize}
\item $X$: big-species lattice point -- $\ell(1, 0, 0)$ 
\item $Y$: small-species ``lattice point'', also called the octahedral hole -- $\ell(0, 1/2, 0)$
\item $A$: half-way between big and small species -- $\ell(0, 1/4, 0)$
\item $B$: half-way between $A$ and $Y$ -- $\ell(0, 3/8, 0)$
\item $C$: half-way between two big-species lattice sites -- $\ell(1/4, 1/4, 0)$
\item $D$: additional site of non-vanishing small-species density, also called the tetrahedral hole -- $\ell(1/4, 1/4, 1/4)$
\end{itemize}
See also \Cref{fig:3d-directions} for a sketch
of the unit cell with the above reference points and the three crystallographic directions $[100], [110] \text{ and } [111]$.
Note, that some figures showing the DCF have two different $y$-axis scales, because in those cases the correlation between the large particles ($LL$) is orders of magnitude larger than the others.

\sfig[0.9\linewidth]{batch_iss_eta55/xs_0.3000/densities_vac_1.0000e-01_DCF_py_comparison_c2_py_comparison.pdf}{Comparison of the PY binary DCF and the FMT crystal DCF for $\nvac =$ \num{0.1}, \hl{$q=0.3$, $x_S = 0.3$ and $\eta=0.5564$} for the ISS phase and the origin positioned at one of the fcc sites, i.e.\ $\rv = \rv_X$ and $d = |\rv - \rv'|$ along the [100] direction.}{fig:iss-c2-py}

First, we inquire about a sensible ``liquid-like'' limit of the crystal DCF. The results of the DCF for the one-component crystal of Ref.~\cite{lin2021direct} and of the SC in \Cref{fig:subst-c2-100} imply that it is very different from the liquid DCF, both in range and magnitude (which is $\sim - 1/\nvac$ for the crystal, whereas the one-component PY-DCF of a dense HS liquid is $-O(10)$).  This suggests to consider a fully minimized crystal with an artificially large vacancy density for a comparison, here we choose $\nvac=0.1$. For such a crystal, the density peaks of the large species are still localized around the fcc sites but show very spread out, anisotropic tails. Similarly, the small species peaks are clearly seen at the interstitial sites but here the spread-out, anisotropic tails cover the whole unit cell. The DCF for this crystal (around reference point $X$) is shown in \Cref{fig:iss-c2-py}, in comparison to the Lebowitz solution for the binary PY closure \cite{lebowitz1964exact}.
We see that the partial DCF including at least one small particle ($LS$, $SS$), are perfectly
in line with the PY result. Only the correlation between large spheres ($LL$) show larger deviations by a factor of two.
Thus one may conclude that in the ISS phase, the direct correlation function of the small species is effectively that of a liquid, despite the fact that $\rho_S(\rv)$ is still very inhomogeneous.%
This is not the case for the large spheres.

\sfig[0.9\linewidth]{combined_DCF_large.pdf}{Equilibrium DCF of the ISS phase with reference point at the  large sphere fcc lattice site along three directions, i.e.\ $\rv=\rv_X$ and $d=|\rv-\rv'|$.}{fig:iss-c2-large}
\sfig[0.9\linewidth]{combined_DCF_small.pdf}{Equilibrium DCF of the ISS phase with reference point at the interstitial lattice site along three directions, i.e.\ $\rv=\rv_Y$ and $d=|\rv-\rv'|$. Note the different $y$-axes for the $LL$ component (left axis) vs. the $LS$ and $SS$ components (right axis).}{fig:iss-c2-small}

Next, we focus on the DCFs at the true vacancy density of the crystal phase. These are shown in \Cref{fig:iss-c2-large}, with the reference point at the big species site ($X$) and in \Cref{fig:iss-c2-small}  at the octahedral interstitial site ($Y$). %
Finally, a three-dimensional plot of the $LL$ component of the DCF, evaluated at all points mentioned before is shown in \Cref{fig:volume-grid}. We do not show the $LS$ and $SS$ components as they are very close to being radially symmetric.

\sfig[1\linewidth]{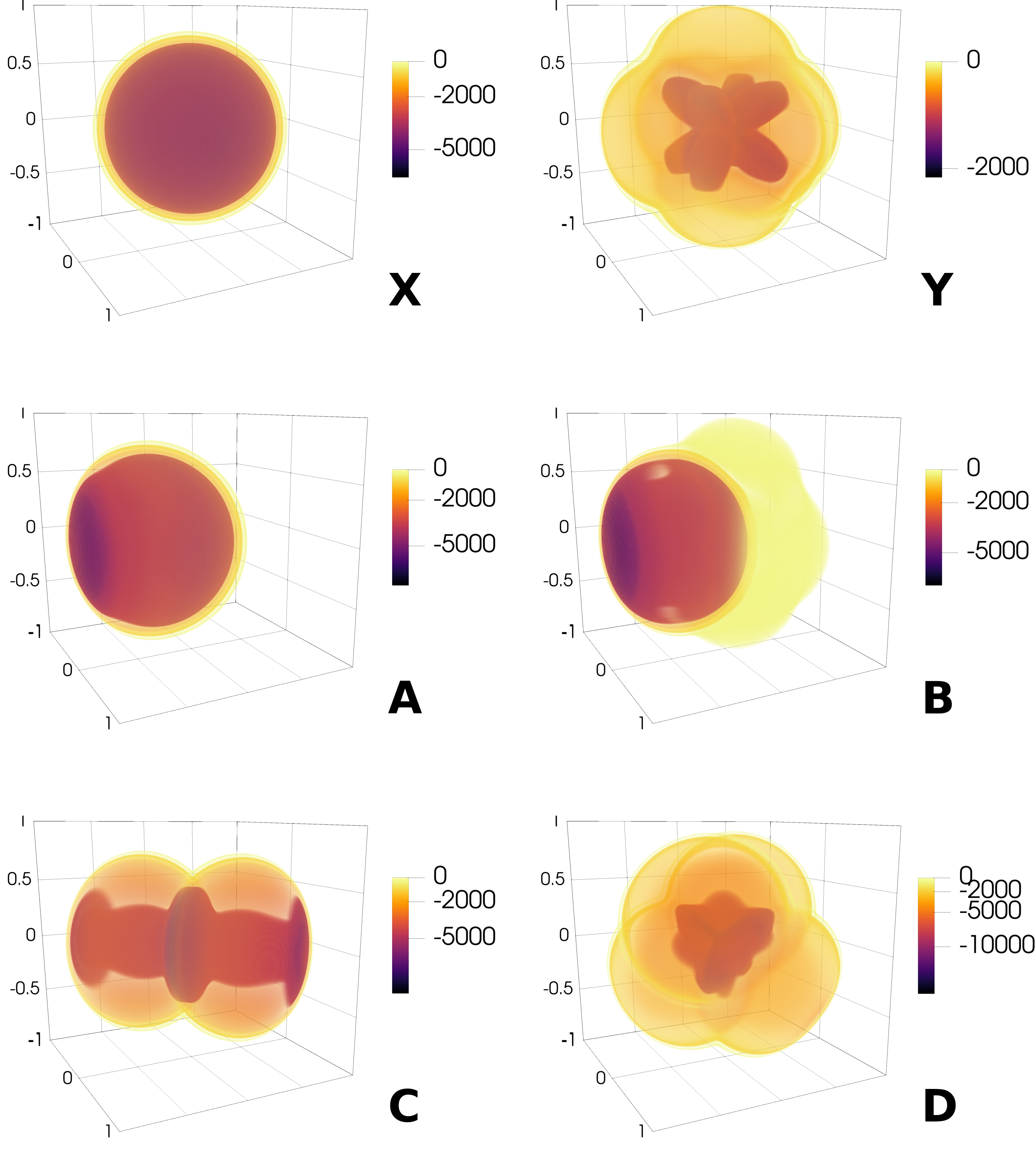}{Volumetric plot of the DCF for the $LL$-component evaluated at different reference points $\rv_i, i \in \{X, Y, \ldots\}$. For a description of the points see the main text and \Cref{fig:3d-directions}. The 3D coordinate system is the one for $\Delta \rv=\rv'-\rv_i$, axes are in units of $\sigma_L$.}{fig:volume-grid}

We make the following observations:
(i) Highly localized density profiles lead to reduced range of the DCF, when compared to the liquid regime. While this was already observed in the single-component case, this becomes clearer for the ISS system. Here the small species DCF is of very large magnitude (much larger than for liquids) but still has the characteristic range given by $\sigma_i/2 + \sigma_j/2$, i.e.\ $0.3\,\sigma_L$ for the $SS$ component and $0.65\,\sigma_L$ for the $LS$ component. The big species on the other hand has a smaller range \hl{(compared to the liquid case or the maximal allowed range according to FMT)} of about $0.8 < 1.0$ (in units of $\sigma_L$).

(ii) All direct correlation functions for the ISS phase differ qualitatively from those we saw in the single-component or SC case. There is no characteristic dip and the functions are much more anisotropic, which is expected since the crystal structure is also more complicated. An exception to this is the lattice point X. Here the DCF is relatively isotropic, for all components. This is due to the fact that X is a high symmetry point and considering the short range of the correlation, rather similar to a pure fcc lattice as seen in the single-component case.

The interpretation of the direct correlation function is more involved than the usual pair correlation function. It is useful to first consider a one-component system and
the first-order DCF $c^{(1)}(\rv)=-\beta \delta F_{\text{exc}}[\rho]/\delta \rho(\rv)$, which for hard spheres is $\ln p_\text{ins}(\rv)$ with 
$p_\text{ins}(\rv)$ being the (inhomogeneous) insertion probability of a hard sphere into an equilibrium system with density distribution $\rho(\rv)$.
\hl{This follows from the potential distribution theorem or alternatively the Widom insertion rule\cite{frenkel2023understanding}, which
states that
$
\cone = \ln \langle e^{-\beta \Delta U(\mathbf{r})} \rangle,
$
where $\Delta U$ is the change in energy when a particle is inserted at $\rv$. Consequently for hard spheres the expectation value becomes just the
insertion probability.
}
Since $\ctwo(\rv,\rv')=\delta c^{(1)}(\rv) / \delta \rho(\rv') $, it describes the change of $\ln p_\text{ins}(\rv)$ upon a slight density change
at $\rv'$. For $\nvac \to 0$,  $p_\text{ins}(\rv)$ at a lattice site should be $\sim \nvac$, as this is the probability of finding the lattice site empty. A positive density change at a second point $\rv'$, with $|\rv-\rv'|<\sigma$, corresponds to having an additional hard sphere at $\rv'$ in some ensemble members (which is only possible if the site $\rv$ is vacant for this member).
\hl{
The insertion of a particle at $\rv'$ lowers the probablity of insertion at $\rv$, i.e.}\ $\delta p_\text{ins}(\mathbf{r}) \sim - \delta \rho(\mathbf{r}')$.
\hl{
Thus, that additional hard sphere reduces
$\nvac$ and one finds $\ctwo(\rv,\rv')=\delta c^{(1)}(\rv) / \delta \rho(\rv') \sim \frac{1}{p_\text{ins}} \partial p_\text{ins}(\rv)/\partial \rho(\rv') = - 1/\nvac $.
}This argument explains the magnitude of the one-component DCF but not its reduced range which is a  subtle effect of the finite width of the density profile around a lattice site. For the two-component solids the argument remains valid if the reference point is the fcc lattice site since the insertion probability at this site is $\sim \nvac$ both for large and small spheres. Therefore all components of the DCF have the magnitude $1/\nvac$ for this reference point. A difference between large and small sphere correlations arises, if the reference point is the interstitial point $Y$ for the ISS. Here, the insertion probability for a large sphere is still $ \sim \nvac$ (vacant nearby lattice sites), but for small spheres it is of the order $1/2-X_S$ (fraction of empty interstitial sites), and consequently all DCF components with small spheres will be of smaller absolute magnitude.

The above considerations motivate a qualitative picture of the $LL$ component of the DCF (for arbitrary reference points) as a superposition of short-ranged and nearly isotropic base functions with reference point at an fcc lattice site. The base function, which is different for different lattice points, reflects the argument that the insertion of a large sphere at an arbitrary reference point needs a vacant site at a nearby fcc lattice point. This picture is exemplified in \Cref{fig:volume-grid} where the DCF is shown as a 3D plot for six different reference points. For the reference point $X$, i.e. the fcc lattice point, the DCF is the base function itself, and neighboring lattice points do not contribute with base functions since they are too far away.
On the other hand, at reference points $Y$, $C$ and $D$ we can clearly see how nearest neighbors shape the form of the DCF.
For the point $Y$ there are six nearest neighbors whose contributions add up, at the point $C$ there are two and at the point $D$ there are four nearest neighbors arranged tetrahedrally. For the reference points $A$ and $B$ (on the line connecting an fcc point and the interstitial point $Y$), the main contribution to the DCF comes from the base function around this fcc point, although for $B$ (closer to $Y$) there is some small contribution seen from the fcc neighbor sites of $Y$.

\hl{
The different shape of $\ctwo_{LL}$ for the ISS and SC phase (cf. \Cref{fig:subst-c2-100,fig:iss-c2-large})  may seem susprising,
since the density distribution for the large species $\rho_L$ is similiar in both cases. However, it is imporant to note that according to \Cref{eq:c2-full}, $\ctwo_{LL}$ is
determined by the Hessian of the free energy with respect to the weighted densities, and the weighted densities contain contributions from both species.
Since in the SC phase the small particles are located on the same lattice as the large ones (and their radius is also similar), the weighted density at these
points is similar to the single-component case. In the ISS phase the small
spheres are concentrated around the octahedral and tetrahedral holes, with neglible contributions in regions connecting the sites. This modifies the weighted densities at the large-particle lattice sites, even though the large-
particle density remains similar.
}

\subsection{``Diffusion pathway'' between octahedral and tetrahedral holes for small spheres in the ISS}

\sfig[.9\linewidth]{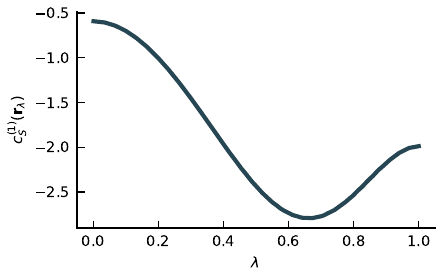}{Value of $\cone_S(\rv)$ on a linear path going from the octahedral hole $\rv_Y$ to the tetrahedral hole $\rv_D$.}{fig:path-c1}

Previously we discussed that the small-species density is less localized in the unit cell than that of the big species. While there are still points with particularly high particle density, i.e.\, the octahedral and tetrahedral holes, the space in between them also has a non-vanishing probability density. These connecting areas can be interpreted as tunnels that allow hopping or diffusion of the small species between the preferred ``lattice points''. To better understand the energy landscape between these point we show the value of $\cone_S(\rv )$ on a linear path from the octahedral hole $Y$ to the tetrahedral hole $D$, parametrized by the variable $\lambda$, where $\rv(\lambda=0)=\rv_Y$ and $\rv(\lambda=1) = \rv_D$ in \Cref{fig:path-c1}. It is the negative excess free energy cost of inserting a small sphere on the path (and as discussed before, this corresponds to $\ln p_\text{ins}(\rv)$), and it is seen to exhibit a moderate barrier of $2 k_\text{B} T$ for crossing from $Y$ to $D$. Thus we expect small spheres in the ISS to be very mobile. Our free energy profile qualitatively agrees with the profiles from ISS simulations \cite{vandermeer2017interstitials} which were taken at a higher packing fraction and resulted in somewhat larger barrier heights.

\section{Discussion and Outlook}
\label{sec:discussion}
Using fundamental measure theory,
we have investigated the equilibrium structure of two colloidal binary crystal phases, i.e.\, the substitutional crystal and the interstitial solid solution. Due to the nature of the free (non-parameterized) DFT minimization we account for different types of defects (vacancy, interstitial and substitutional defects), going beyond the ideal crystal. The 3D resolved ensemble-averaged density profiles of the unit cell show that in the substitutional crystal, the large and small sphere profiles are very similar to the one-component solid (being narrow Gaussians around the fcc lattice sites), while for the interstitial solid solution the small sphere profile is different in being delocalized over the whole unit cell, with a maximum at the octahedral interstitial site and a secondary maximum at the tetrahedral interstitial site.
We computed the species-resolved direct correlation function inside the unit cell around several reference points. All components of the direct correlation function in the substitutional crystal and the large--large component in the interstitial solid solution show the characteristic magnitude $-O(1/\nvac)$ already found in the one-component crystal, much larger than that of a high-density liquid. For arbitrary reference points, these components of the direct correlation function can be understood as resulting from the superposition of fcc lattice point centered base functions which are approximately isotropic.
Further, we note the reduced range of correlation, compared to the liquid case, when the density peaks become highly localized.
These results for the direct correlation functions in the binary crystal case further add to the portfolio of the basic two-point correlations for the phases of simple fluids. 

Furthermore, we expect these results to be useful for the evaluation of generalized elastic constants, in analogy to the calculations of Ref.~\cite{lin2021direct} for the single-component case.

\emph{Acknowledgments:} Funded by the Deutsche Forschungsgemeinschaft (DFG, German Research Foundation) – Projektnummer 401742484 (WEAVE project ``Binary complex colloidal crystals: Role of substitutional disorder and stacking faults''). We thank our project partners Matthias Fuchs, Gerhard Kahl and Rudolf Haussmann for useful discussions and Matthias Fuchs additionally for a critical reading of the manuscript.  

\nocite{*}
\bibliography{binhsref}

@article{rosenfeld1989,
  author    = {Rosenfeld, Yaakov},
  title     = {Free-energy model for the inhomogeneous hard-sphere fluid mixture},
  journal   = {Phys. Rev. Lett.},
  volume    = {63},
  number    = {9},
  pages     = {980--983},
  year      = {1989},
  publisher = {American Physical Society},
  doi       = {10.1103/PhysRevLett.63.980}
}

@article{lebowitz1964exact,
  author    = {Lebowitz, Joel L.},
  title     = {Exact solution of generalized {Percus--Yevick} equation for a mixture of hard spheres},
  journal   = {Phys. Rev.},
  volume    = {133},
  number    = {4A},
  pages     = {A895--A899},
  year      = {1964},
  publisher = {American Physical Society},
  doi       = {10.1103/PhysRev.133.A895}
}

@article{filion2011self,
  author    = {Filion, Laura and Hermes, Michiel and Ni, Ran and Vermolen, E. C. M. and Kuijk, Anke and Christova, C. G. and Stiefelhagen, J. C. P. and Vissers, Teun and van Blaaderen, Alfons and Dijkstra, Marjolein},
  title     = {Self-assembly of a colloidal interstitial solid with tunable sublattice doping},
  journal   = {Phys. Rev. Lett.},
  volume    = {107},
  number    = {16},
  pages     = {168302},
  year      = {2011},
  publisher = {American Physical Society},
  doi       = {10.1103/PhysRevLett.107.168302}
}

@article{royall2024colloidal,
  author    = {Royall, C. Patrick and Charbonneau, Patrick and Dijkstra, Marjolein and Russo, John and Smallenburg, Frank and Speck, Thomas and Valeriani, Chantal},
  title     = {Colloidal hard spheres: Triumphs, challenges, and mysteries},
  journal   = {Rev. Mod. Phys.},
  volume    = {96},
  number    = {4},
  pages     = {045003},
  year      = {2024},
  publisher = {American Physical Society},
  doi       = {10.1103/RevModPhys.96.045003}
}

@article{hansen2006density,
  author    = {Hansen-Goos, Hendrik and Roth, Roland},
  title     = {Density functional theory for hard-sphere mixtures: The {White Bear} version {Mark II}},
  journal   = {J. Phys.: Condens. Matter},
  volume    = {18},
  number    = {37},
  pages     = {8413--8425},
  year      = {2006},
  publisher = {IOP Publishing},
  doi       = {10.1088/0953-8984/18/37/002}
}

@article{lin2021direct,
  author    = {Lin, S.-C. and Oettel, Martin and H{\"a}ring, Johannes M. and Haussmann, Rudolf and Fuchs, Matthias and Kahl, Gerhard},
  title     = {Direct correlation function of a crystalline solid},
  journal   = {Phys. Rev. Lett.},
  volume    = {127},
  number    = {8},
  pages     = {085501},
  year      = {2021},
  publisher = {American Physical Society},
  doi       = {10.1103/PhysRevLett.127.085501}
}

@article{oettel2010free,
  author    = {Oettel, Martin and G{\"o}rig, S. and H{\"a}rtel, A. and L{\"o}wen, Hartmut and Radu, Marc and Schilling, Tanja},
  title     = {Free energies, vacancy concentrations, and density distribution anisotropies in hard-sphere crystals: A combined density functional and simulation study},
  journal   = {Phys. Rev. E},
  volume    = {82},
  number    = {5},
  pages     = {051404},
  year      = {2010},
  publisher = {American Physical Society},
  doi       = {10.1103/PhysRevE.82.051404}
}

@book{hansen2013theory,
  author    = {Hansen, Jean-Pierre and McDonald, Ian Ranald},
  title     = {Theory of Simple Liquids: With Applications to Soft Matter},
  edition   = {4},
  year      = {2013},
  publisher = {Academic Press},
  address   = {Oxford},
  doi       = {10.1016/C2010-0-66723-X},
  isbn      = {978-0-12-387032-2}
}

@book{frenkel2023understanding,
  author = {Daan Frenkel and Berend Smit},
  title = {Understanding Molecular Simulation},
  edition   = {3},
  year      = {2023},
  publisher = {Academic Press},
  doi       = {10.1016/C2009-0-63921-0},
  isbn      = {978-0-323-90292-2}
}

@article{evans1979nature,
  author    = {Evans, Robert},
  title     = {The nature of the liquid-vapour interface and other topics in the statistical mechanics of non-uniform, classical fluids},
  journal   = {Adv. Phys.},
  volume    = {28},
  number    = {2},
  pages     = {143--200},
  year      = {1979},
  publisher = {Taylor \& Francis},
  doi       = {10.1080/00018737900101365}
}

@article{roth2010fundamental,
  author    = {Roth, Roland},
  title     = {Fundamental measure theory for hard-sphere mixtures: A review},
  journal   = {J. Phys.: Condens. Matter},
  volume    = {22},
  number    = {6},
  pages     = {063102},
  year      = {2010},
  publisher = {IOP Publishing},
  doi       = {10.1088/0953-8984/22/6/063102}
}

@article{curtin1986density,
  author    = {Curtin, W. A. and Ashcroft, N. W.},
  title     = {Density-functional theory and freezing of simple liquids},
  journal   = {Phys. Rev. Lett.},
  volume    = {56},
  number    = {26},
  pages     = {2775--2778},
  year      = {1986},
  publisher = {American Physical Society},
  doi       = {10.1103/PhysRevLett.56.2775}
}

@article{lutsko2020classical,
  title={Classical density-functional theory applied to the solid state},
  author={Lutsko, James F and Schoonen, C{\'e}dric},
  journal={Phys. Rev. E},
  volume={102},
  number={6},
  pages={062136},
  year={2020},
    publisher = {American Physical Society},
  doi= {10.1103/PhysRevE.102.062136}
}

@article{lutsko2020explicitly,
  title={Explicitly stable fundamental-measure-theory models for classical density functional theory},
  author={Lutsko, James F},
  journal={Phys. Rev. E},
  volume={102},
  number={6},
  pages={062137},
  year={2020},
    publisher = {American Physical Society},
  doi = {10.1103/PhysRevE.102.062137}
}

@article{haertel2012prl,
  title={Tension and stiffness of the hard sphere crystal-fluid interface},
  author={H{\"a}rtel, Andreas and Oettel, Martin and Rozas, Roberto E and Egelhaaf, Stefan U and Horbach, J{\"u}rgen and L{\"o}wen, Hartmut},
  journal={Phys. Rev. Lett.},
  volume={108},
  number={22},
  pages={226101},
  year={2012},
  publisher={APS},
  doi = {10.1103/PhysRevLett.108.226101}
}

@article{oettel2012comparison,
  title={Description of hard-sphere crystals and crystal-fluid interfaces: A comparison between density functional approaches and a phase-field crystal model},
  author={Oettel, M and Dorosz, S and Berghoff, M and Nestler, B and Schilling, Tanja},
  journal={Phys. Rev. E},
  volume={86},
  number={2},
  pages={021404},
  year={2012},
  publisher={APS},
  doi={10.1103/PhysRevE.86.021404}
}

@article{oettel2012modes,
  title={Mode expansion for the density profiles of crystal--fluid interfaces: hard spheres as a test case},
  author={Oettel, Martin},
  journal={J. Phys.: Condens. Matter},
  volume={24},
  number={46},
  pages={464124},
  year={2012},
  publisher={IOP Publishing},
  doi={10.1088/0953-8984/24/46/464124}
}

@article{kwak2008simvac,
  title={Characterization of mono- and divacancy in fcc and hcp hard-sphere crystals},
  author={Kwak, Sang Kyu and Cahyana, Yenni and Singh, Jayant K},
  journal={J. Chem. Phys.},
  volume={128},
  number={13},
  pages = {134505},
  year={2008},
  publisher={AIP Publishing},
  doi = {10.1063/1.2889924}
}

@article{hansen-goos2015santos,
  title={Fundamental measure theory for the inhomogeneous hard-sphere system based on Santos' consistent free energy},
  author={Hansen-Goos, Hendrik and Mortazavifar, Mostafa and Oettel, Martin and Roth, Roland},
  journal={Phys. Rev. E},
  volume={91},
  number={5},
  pages={052121},
  year={2015},
  doi={10.1103/PhysRevE.91.052121},
  publisher={APS}
}

@article{schoonen2022prl,
  title={Crystal polymorphism induced by surface tension},
  author={Schoonen, C{\'e}dric and Lutsko, James F},
  journal={Phys. Rev. Lett.},
  volume={129},
  number={24},
  pages={246101},
  year={2022},
  publisher={APS},
  doi={10.1103/PhysRevLett.129.246101}  
}

@book{santosbook,
  title={A Concise Course on the Theory of Classical Liquids},
  author={Santos, Andr{\'e}s},
  year={2016},
  publisher={Springer},
  doi={10.1007/978-3-319-29668-5}
}

@article{kranendonk1991coexistence,
  author  = {Kranendonk, W. G. T. and Frenkel, D.},
  title   = {Computer simulation of solid-liquid coexistence in binary hard sphere mixtures},
  journal = {Mol. Phys.},
  volume  = {72},
  number  = {3},
  pages   = {679--697},
  year    = {1991},
  doi     = {10.1080/00268979100100501}
}

@article{eldridge1993superlattice,
  author  = {Eldridge, M. D. and Madden, P. A. and Frenkel, D.},
  title   = {Entropy-driven formation of a superlattice in a hard-sphere binary mixture},
  journal = {Nature},
  volume  = {365},
  pages   = {35--37},
  year    = {1993},
  doi     = {10.1038/365035a0}
}

@article{filion2009prediction,
  author  = {Filion, Laura and Dijkstra, Marjolein},
  title   = {Prediction of binary hard-sphere crystal structures},
  journal = {Phys. Rev. E},
  volume  = {79},
  pages   = {046714},
  year    = {2009},
  doi     = {10.1103/PhysRevE.79.046714}
}

@article{hynninen2009stability,
  author  = {Hynninen, A.-P. and Filion, L. and Dijkstra, M.},
  title   = {Stability of {$LS$} and {$LS_2$} crystal structures in binary mixtures of hard and charged spheres},
  journal = {J. Chem. Phys.},
  volume  = {131},
  pages   = {064902},
  year    = {2009},
  doi     = {10.1063/1.3182724}
}

@article{vandermeer2017interstitials,
  author  = {van der Meer, Berend and Lathouwers, Emma and Smallenburg, Frank and Filion, Laura},
  title   = {Diffusion and interactions of interstitials in hard-sphere interstitial solid solutions},
  journal = {J. Chem. Phys.},
  volume  = {147},
  pages   = {234903},
  year    = {2017},
  doi     = {10.1063/1.5003905}
}

@phdthesis{filion2011thesis,
 author =  {Filion, Laura},
  title  = {Self-assembly in colloidal hard-sphere systems},
  school = {Utrecht University},
  year   = {2011},
  type   = {{Ph.\,D.} thesis},
  url    = {http://hdl.handle.net/1874/192603},
}
\clearpage
\appendix

\section{WBII tensor functional and numerical details}

The excess free energy functional within the Fundamental Measure Theory (FMT) framework is given by:
\begin{equation}
    \beta F_{\text{exc}}[\{\rho_i\}] = \int d\mathbf{r} \, \Phi(\{n_\alpha(\mathbf{r})\}),
    \label{eq:Phi}
\end{equation}
where the weighted densities $n_\alpha(\mathbf{r})$ are calculated via convolution of the density profiles $\rho_i(\mathbf{r})$ of species $i$ with the geometric weight functions $w_\alpha^{(i)}$:
\begin{equation}
    n_\alpha(\mathbf{r}) = \sum_i \int d\mathbf{r}' \, \rho_i(\mathbf{r}') w_\alpha^{(i)}(\mathbf{r} - \mathbf{r}').
\end{equation}

In order to get equilibrium density profiles we solve the Euler--Lagrange (EL) equation that results from minimizing the grand potential:
\begin{equation}
  \label{eq:el}
  \rho_{\text{eq},i}(\rv) = \exp \left( \beta \mu_i + \cone_i[\rho_{\text{eq},i}](\rv) \right).
\end{equation}
For this is it is necessary to evaluate the (first) direct correlation function
\begin{equation}
  \label{eq:DCF1}
  \cone_i(\rv) = -\fdv{\beta F_\text{exc}[\rho]}{\rho_i(\rv)},
\end{equation}
which we do by implementing the calculation of $F_\text{exc}$ as a fully differentiable expression in one of the usual machine-learning frameworks.
The density field in one unit cell is discretized on a (periodic) grid with $128 \times 128 \times 128$ points and initialized with Gaussian peaks at the appropriate lattice sites and realistic widths.
We then solve the EL equation iteratively using a mix of Picard and DIIS (direct inversion in the iterative subspace) steps. During these iterations the chemical potentials $\mu_i$ are chosen such that the mean particle number inside the unit cell is fixed. This means that during the solution procedure $\mu_i$ vary, but ultimately converge
to the true value the closer we get to the fixed point. We stop the iteration once the error of the $j$th iteration $\rho_i^{(j)}(\rv)$ (summed squared mean difference between two iterations)
becomes smaller than $\num{1e-5}$. Computing an equilibrium density profile on a GPU typically requires approximately 10 minutes. \hl{The unit cell geometry and length as defined in \Cref{eq:length-def} remained constant throughout the minimization.}

The computation of $\ctwo_{ij}(\rv, \rv')$ is similarly done by automatic differentiation, however the first argument needs to be fixed due to memory limitations.
\subsection*{Weight Functions}
For hard spheres of radius $R_i$, the scalar weight functions are
\begin{subequations}
\begin{align}
    w_3^{(i)}(\mathbf{r}) &= \Theta(R_i - |\mathbf{r}|), \\
    w_2^{(i)}(\mathbf{r}) &= \delta(R_i - |\mathbf{r}|), \\
    w_1^{(i)}(\mathbf{r}) &= \frac{w_2^{(i)}(\mathbf{r})}{4\pi R_i}, \\
    w_0^{(i)}(\mathbf{r}) &= \frac{w_2^{(i)}(\mathbf{r})}{4\pi R_i^2}.
\end{align}
\end{subequations}
The vector weight functions are
\begin{subequations}
\begin{align}
    \mathbf{w}_2^{(i)}(\mathbf{r}) &= \frac{\mathbf{r}}{|\mathbf{r}|}\,\delta(R_i - |\mathbf{r}|), \\
    \mathbf{w}_1^{(i)}(\mathbf{r}) &= \frac{\mathbf{w}_2^{(i)}(\mathbf{r})}{4\pi R_i}.
\end{align}
\end{subequations}
The tensor weight functions (traceless form) are given by
\begin{equation}
    \mathbf{w}_{T}^{(i)}(\mathbf{r})
    = \left(\frac{\mathbf{r}\mathbf{r}}{|\mathbf{r}|^2} - \frac{1}{3}\mathbf{I}\right)\delta(R_i-|\mathbf{r}|),
\end{equation}
where $\mathbf{I}$ is the identity matrix and $\mathbf{r}\mathbf{r}$ is the dyadic product.

\subsection*{Free Energy Density}
The White Bear II tensor functional (traceless tensor version) is most conveniently written as
\begin{multline}
    \Phi_{\mathrm{WBII}} =
    -n_0 \ln(1-n_3)
    + \frac{1+\tfrac{1}{3}\phi_2(n_3)}{1-n_3}
      \left(n_1 n_2 - \mathbf{n}_1 \cdot \mathbf{n}_2\right) \\
    + \frac{1-\tfrac{1}{3}\phi_3(n_3)}{24\pi(1-n_3)^2}
      \left[
        n_2^3 - 3n_2\,\mathbf{n}_2\cdot\mathbf{n}_2
        + 9\left(\mathbf{n}_2\cdot\mathbf{n}_{T}\cdot\mathbf{n}_2
        - \frac{1}{2}\operatorname{Tr}\!\left(\mathbf{n}_{T}^3\right)\right)
      \right].
\end{multline}
Here $\mathbf{n}_{T}$ is the weighted density generated by $\mathbf{w}_{T}^{(i)}$.

The WBII scalar functions are
\begin{subequations}
\begin{align}
    \phi_2(n_3) &= \frac{1}{n_3}\left[2n_3 - n_3^2 + 2(1-n_3)\ln(1-n_3)\right], \\
    \phi_3(n_3) &= \frac{1}{n_3^2}\left[2n_3 - 3n_3^2 + 2n_3^3 + 2(1-n_3)^2\ln(1-n_3)\right].
\end{align}
\end{subequations}

\end{document}